\documentclass[a4paper, oneside, twocolumn, notitlepage, 10pt]{extarticle_ecoc}
\usepackage{ecoc}
\usepackage{upgreek}
\usepackage{multirow}
\usepackage[position=top]{subfig}
\usepackage{tikz,pgfplots,subfig, filecontents}
\usepackage{graphicx}
\usepackage{multicol}
\usepackage{colortbl}%
  \newcommand{\myrowcolour}{\rowcolor[gray]{0.925}}
\usepackage{longtable}
\newcommand{\mytab}{
\begin{tabular}{c c c c c c c c c c c}
\hline
Type    & \begin{tabular}[c]{@{}c@{}}Latency\\ ($\mu s$)\end{tabular} & \begin{tabular}[c]{@{}c@{}}Clock\\ Frequency\\ (MHz)\end{tabular} & BRAM & SRL & \begin{tabular}[c]{@{}c@{}}DSP \\ Slices\end{tabular} & LUT & FF  & \begin{tabular}[c]{@{}c@{}}Utilization (\%)\\ $[$ DSP/LUT/FF $]$\end{tabular}& \begin{tabular}[c]{@{}c@{}}Throughput\\ ($Gbits/s$)\end{tabular}  & \begin{tabular}[c]{@{}c@{}}$N^{o}$ FPGAs\\ for 400G\end{tabular} \\ \hline\hline
\myrowcolour
biLSTM & 33.4                                                    & 270                                                               & 164  & 109 & 1260                                                   & 113532 & 224386 & \textcolor{red}{64.0}/12.6/12.5 & 66                                                            & 4.8                                                        \\ 

CNN & 19.9                                                    & 244                                                               & 0  & 125 & 582    & 118477                                                & 379829 & \textcolor{red}{29.6}/13.2/21.1 & 60                                                            & 2.6                                                    
\\
\myrowcolour%
CDC     & 1.1                                                    & 524                                                               & 0  & 1 & 1072 & 10441 & 5640 & \textcolor{red}{54.5}/1.2/0.3 & 127                                                            & 2.1                                                        \\ \hline

\end{tabular}
}

\hypersetup{bookmarksdepth=-2}
\begin{document}
\selectlanguage{english}    


\title{Towards FPGA Implementation of Neural Network-Based  Nonlinearity Mitigation Equalizers in Coherent Optical Transmission Systems }%


\author{
    Pedro J. Freire\textsuperscript{(1)}, Michael Anderson\textsuperscript{(1)}, Bernhard Spinnler\textsuperscript{(2)}, Thomas Bex\textsuperscript{(2)}, Jaroslaw E. Prilepsky\textsuperscript{(1)}, \\ Tobias A. Eriksson \textsuperscript{(3)}, Nelson Costa\textsuperscript{(4)}, Wolfgang Schairer\textsuperscript{(2)}, Michaela Blott\textsuperscript{(5)}, Antonio Napoli\textsuperscript{(2)}, \\ Sergei K. Turitsyn\textsuperscript{(1)}
}
\maketitle                  

\vspace{-1.5mm}
\begin{strip}
 \begin{author_descr}

   \textsuperscript{(1)}~Aston University, United Kingdom
   \textcolor{blue}{\uline{p.freiredecarvalhosourza@aston.ac.uk}} 
   \textsuperscript{(2)}~Infinera, Munich, Germany
   \textsuperscript{(3)}~Infinera, Stockholm,  Sweden\textsuperscript{(4)}~Infinera, Lisbon, Portugal
   \textsuperscript{(5)}~AMD/Xilinx Inc, Dublin, Ireland

 \end{author_descr}
\end{strip}

\setstretch{1.1}

\vspace{-1.5mm}
\begin{strip}
\begin{ecoc_abstract}
    For the first time, recurrent and feedforward neural network-based equalizers for nonlinearity compensation are implemented in an FPGA, with a level of complexity comparable to that of a dispersion equalizer. We demonstrate that the NN-based equalizers can outperform a 1-step-per-span DBP. 
\end{ecoc_abstract}
\end{strip}


\section{Introduction}
\vspace{-1mm}
Upcoming technologies and innovations (6G, high-speed access networks, etc.) will push optical networks to meet even more stringent traffic requirements. However, the Kerr effect limits the information rates that optical fiber communication systems can achieve\cite{Cartledge:17,7389319,9748613}. Machine learning-based solutions have lately been touted as a possible approach to mitigate the impact of this fiber transmission effect~\cite{musumeci2018overview,khan2019optical,9585630,9239878,freire2021}. In particular, neural network (NN) algorithms have already been shown to outperform traditional digital backpropagation (DBP) solutions while requiring less computational complexity\cite{9240033,ming2021ultralow}. However, research has often been carried out with simulated datasets only, with the complexity comparisons being made in terms of number of real multiplications.

In this paper, we make an important step forward in assessing the viability of NN-based equalizers for industrial applications by benchmarking i) their performance versus the 1-step-per-span (StpS) DBP using 2.3~samples/symbol (Sa/symbol) in experiments and ii) their computational complexity by comparing an FPGA implementation against the full electronic chromatic dispersion compensation (CDC) block (used, e.g., in standard DSP chain~\cite{kuschnerov2009dsp}), which needs far less resources than 1~StpS DBP. In addition, to the best of our knowledge, for the first time, we present the FPGA implementation of a NN-based equalizer that employs the bidirectional recurrent layer with LSTM cells (biLSTM). By transmitting a 34 GBd single-channel, dual-polarization (SC-DP) 16QAM  signal over 17$\times$70 km of large-effective area fiber (LEAF) (both simulated and experimental), we report $\approx\!1.7$ dB Q-factor improvement over a standard DSP chain while requiring only $\approx\!2.5 \times$ more FPGA resources than the CDC block implementation to achieve 400G transmission.
\vspace{-1mm}
\section{Neural Network Design on FPGAs}
\begin{figure*}[!htb]
\centering 
\vspace{-5mm}
  \includegraphics[width=0.95\linewidth]{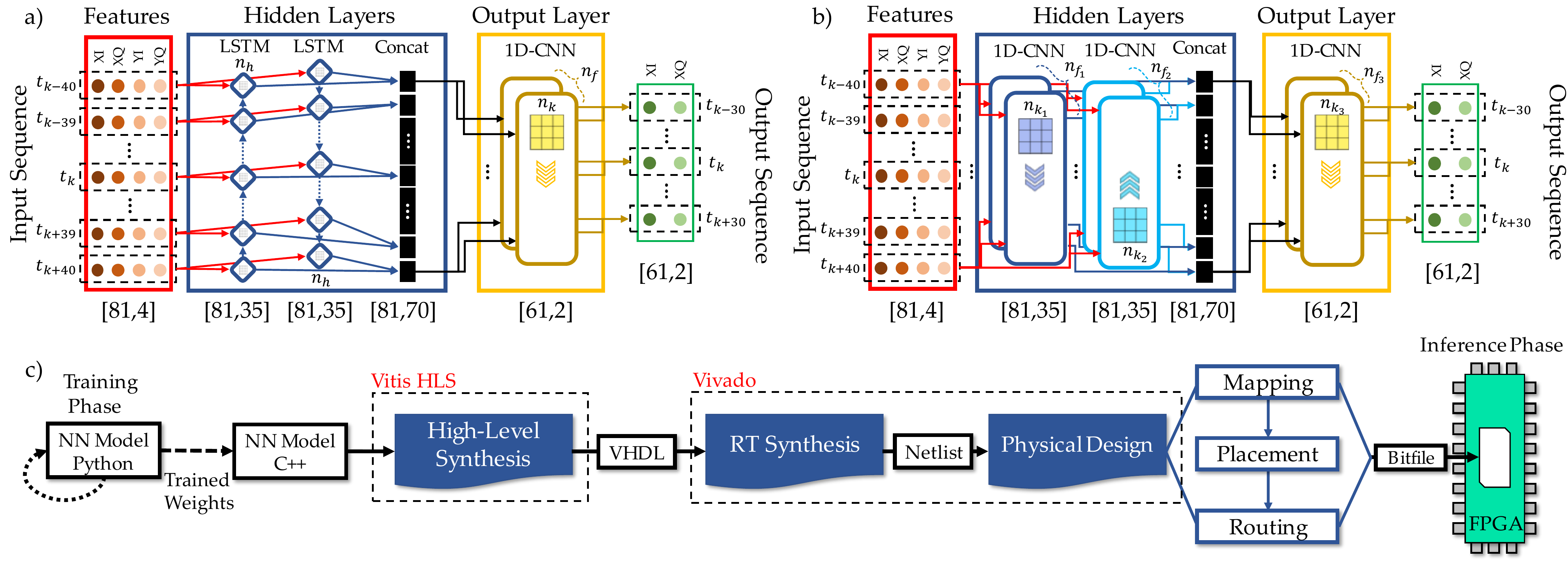}
  \caption{Recurrent equalizer using a) the bidirectional LSTM layer; b) the feedforward equalizer using a 1D-convolutional layer. The pipeline process of the FPGA design from the training in Python to the final realization step on the FPGA is depicted in c).}\label{fig:figure1}
\end{figure*}

The two investigated NN architectures are depicted in Fig.~\ref{fig:figure1}~(a) and (b) for the biLSTM-based equalizer and the deep CNN-based equalizer, respectively. The shape of both architectures is similar, but the mathematical operation nature in each case is different: a) consists of a recursive type whereas b) of a feedforward type. A total of 81 symbols are used as input to simultaneously retrieve 61 symbols on the output. In Fig.~\ref{fig:figure1}, the hidden layer in (a) consists of a biLSTM layer with $n_h=35$ hidden units, and in (b) it is made up of a CNN layer with 70 filters ($n_{f_1}=n_{f_2}=35$), with zero padding applied to retain the shape. The output layer in both designs is a convolutional layer with $n_f=2$ filters, a kernel size of $n_k =21$, and without any padding\footnote{Those parameters were chosen based on grid search analysis to meet the hardware constraints, required throughput, and optical performance.}. Both hidden layers used Tanh activation functions, and the output layer includes only a linear function.

Concerning the FPGA realization, as depicted in Fig.~\ref{fig:figure1}~(c), the design can be broken down into three steps: i) training, ii) high-level synthesis (HLS), and iii) hardware synthesis (RTL synthesis + physical design). In step i), we used the standard TensorFlow library to train the NNs and saved their weights in a fixed-point representation (\texttt{int32}) to be used as input to the NN realization. Then, for each TensorFlow layer, we wrote the respective C++ functions. Afterwards, in step ii), we translated the C++ NN architectures into a hardware description language (VHDL in our case) using the Xilinx software Vitis HLS. Finally, in step iii), the bitstream was generated using the Xilinx software Vivado. This is when the place-and-route process is performed and utilization, chip routing, and other time constraints are verified and defined in a final implemented design. Regarding the CDC implementation, we designed a time-domain equalizer with 556 taps in C++ and followed the same design steps ii) and iii) as in the NN implementation. 

\vspace{-1mm}
\section{Experimental and Simulation Setups}

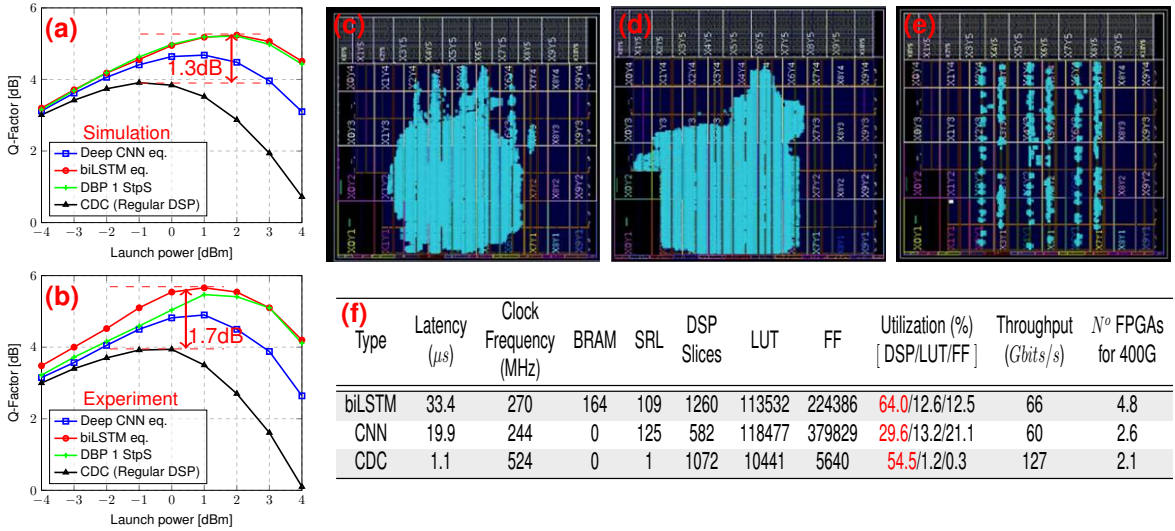
\begin{figure*}
\vspace{-5mm}
    \begin{minipage}[c]{\linewidth}
         \begin{tikzpicture}[scale=0.5]
    \begin{axis} [ylabel={BER}, 
        xlabel={Launch power [dBm]},
        ylabel={Q-Factor [dB]},
        grid=both,  
         ylabel near ticks,
        xmin=-4, xmax=4,
    	xtick={-4, ..., 4},
    	ymin=0, ymax=6,
        legend style={legend pos=south west, legend cell align=left,fill=white, fill opacity=0.6, draw opacity=1,text opacity=1},
    	grid style={dashed}]
        ]
        \addplot[color=blue, mark=square, very thick]     coordinates {
    (-4, 3.12)(-3, 3.62)(-2, 4.06)(-1, 4.41)(0, 4.64)(1,4.68)(2,4.48)(3,3.96)(4,3.1)
    };
    \addlegendentry{Deep CNN eq.};
    
    \addplot[color=red, mark=*, very thick]   
    coordinates {
    (-4, 3.2)(-3, 3.71)(-2, 4.18)(-1, 4.56)(0, 4.95)(1,5.18)(2,5.24)(3,5.06)(4,4.51)
    };
    \addlegendentry{biLSTM eq.};

        \addplot[color=green, mark=+,very thick]    coordinates {
    (-4, 3.16)(-3, 3.69)(-2, 4.19)(-1, 4.63)(0, 4.98)(1,5.19)(2,5.21)(3,4.98)(4,4.43)
    };
    \addlegendentry{DBP 1 StpS};
        \addplot[color=black,mark=triangle, very thick]    coordinates {
    (-4,3.01)(-3, 3.42)(-2, 3.74)(-1, 3.91)(0, 3.84)(1,3.52)(2,2.87)(3,1.94)(4,0.72)
    };
    \addlegendentry{CDC (Regular DSP)};
        \end{axis}
        \node[text width=3cm] at (3.1,5.1) 
    {\textcolor{red}{\textbf{(a)}}};
    \draw[red!80!pink,dashed] (2.6,3.7) -- (5.8,3.7);
    \draw[red!80!pink,dashed] (2.6,5.) -- (5.8,5.);
    \draw[thick, <->,red] (5,3.7) -- +(0,1.3);
    \node[text width=1cm] at (4.3,4.1) 
    {\textcolor{red}{\footnotesize 1.3dB}};
        \node[text width=1cm] at (2.1,2.4) 
    {\textcolor{red}{\scriptsize Simulation}};
    \end{tikzpicture}
    \begin{picture}(100,100)
\put(0,0){ \includegraphics[width=0.228\linewidth, height=0.215\linewidth]{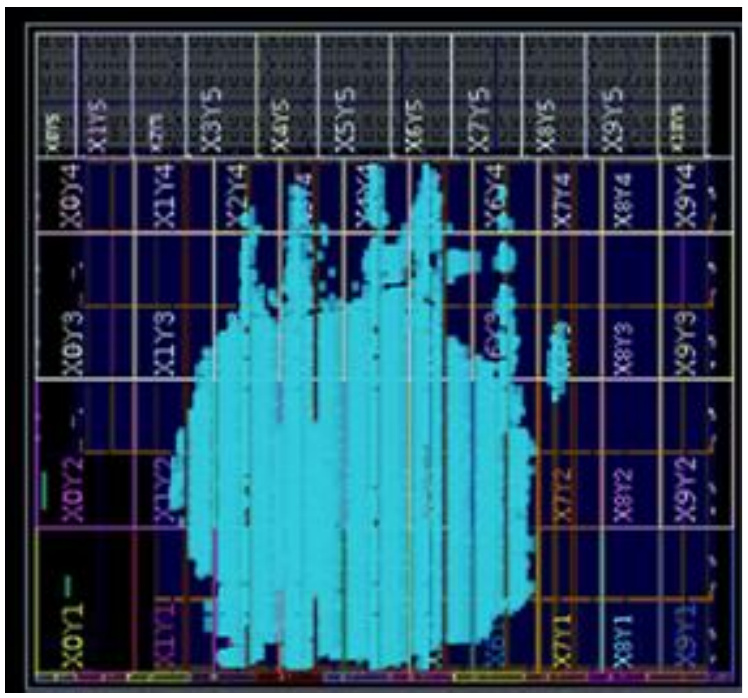}}
\put(6,87){\textcolor{red}{\textbf{(c)}}}
\end{picture}
    \begin{picture}(100,100)
\put(4,0){ \includegraphics[width=0.228\linewidth, height=0.215\linewidth]{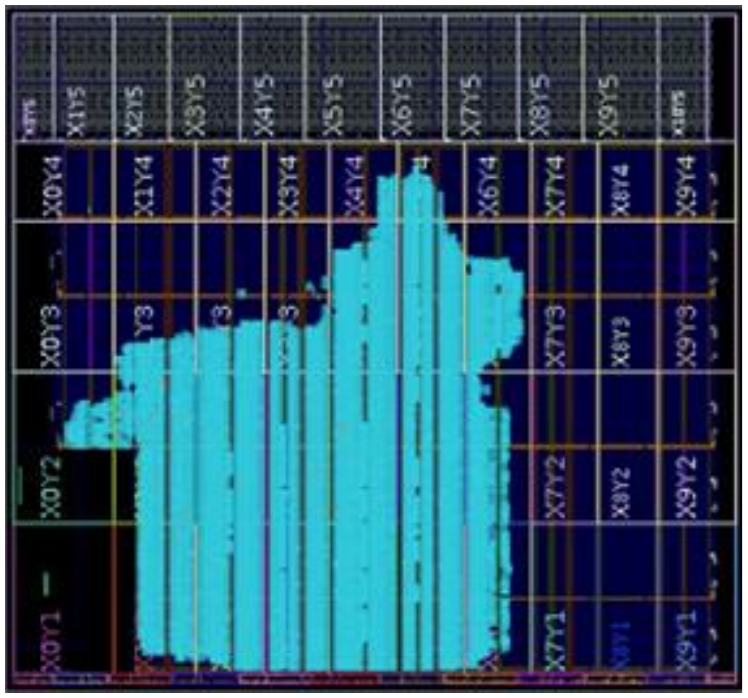}}
\put(10,87){\textcolor{red}{\textbf{(d)}}}
\end{picture}
    \begin{picture}(100,100)
\put(8,0){ \includegraphics[width=0.228\linewidth, height=0.215\linewidth]{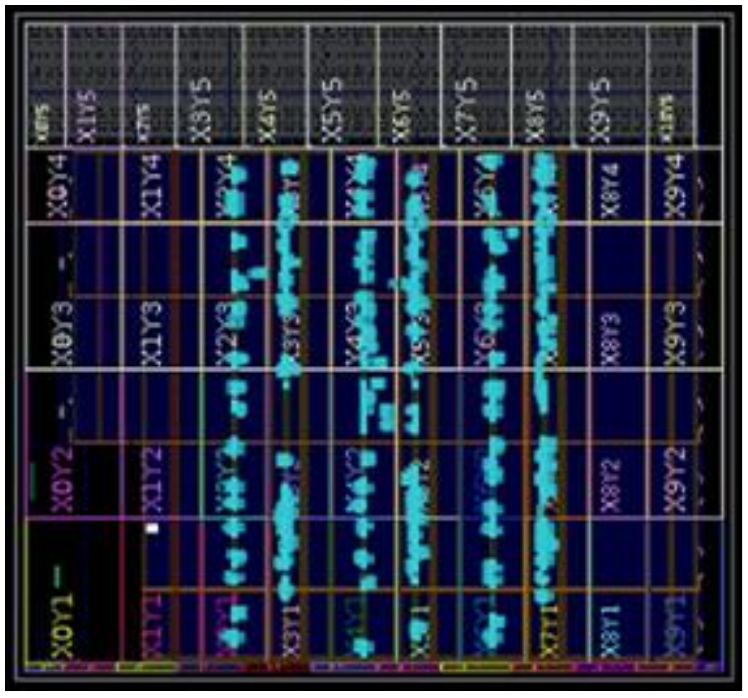}}
\put(14,87){\textcolor{red}{\textbf{(e)}}}
\end{picture}
    \end{minipage}

    \begin{minipage}[c]{\linewidth}
         \begin{tikzpicture}[scale=0.5]
    \begin{axis} [ylabel={BER}, 
        xlabel={Launch power [dBm]},
        ylabel={Q-Factor [dB]},
        grid=both,  
         ylabel near ticks,
        xmin=-4, xmax=4,
    	xtick={-4, ..., 4},
    	ymin=0, ymax=6,
        legend style={legend pos=south west, legend cell align=left,fill=white, fill opacity=0.6, draw opacity=1,text opacity=1},
    	grid style={dashed}]
        ]
        \addplot[color=blue, mark=square, very thick]     coordinates {
    (-4, 3.15)(-3, 3.57)(-2, 4.05)(-1, 4.5)(0, 4.82)(1,4.9)(2,4.5)(3,3.88)(4,2.64)
    };
    \addlegendentry{Deep CNN eq.};
    
    \addplot[color=red, mark=*, very thick]   
    coordinates {
    (-4, 3.48)(-3, 4)(-2, 4.52)(-1, 5.1)(0, 5.54)(1,5.66)(2,5.54)(3,5.1)(4,4.2)
    };
    \addlegendentry{biLSTM eq.};

        \addplot[color=green, mark=+,very thick]    coordinates {
    (-4, 3.22)(-3, 3.72)(-2, 4.16)(-1, 4.59)(0, 5.04)(1,5.47)(2,5.41)(3,5.1)(4,4.11)
    };
    \addlegendentry{DBP 1 StpS};
        \addplot[color=black,mark=triangle, very thick]    coordinates {
    (-4,3)(-3, 3.4)(-2, 3.7)(-1, 3.92)(0, 3.94)(1,3.5)(2,2.7)(3,1.61)(4,0.1)
    };
    \addlegendentry{CDC (Regular DSP)};
        \end{axis}
        \node[text width=3cm] at (3.1,5.1) 
    {\textcolor{red}{\textbf{(b)}}};
    \draw[red!80!pink,dashed] (1.8,3.75) -- (4.8,3.75);
    \draw[red!80!pink,dashed] (1.8,5.4) -- (4.8,5.4);
    \draw[thick, <->,red] (3.8,3.75) -- +(0,1.6);
        \node[text width=1cm] at (4.85,4.1) 
    {\textcolor{red}{\footnotesize 1.7dB}};
            \node[text width=1cm] at (2.1,2.4) 
    {\textcolor{red}{\scriptsize Experiment}};
    \end{tikzpicture}
  \begin{picture}(100,100)
\put(0,0){ \raisebox{3.1cm}{\subfloat{
 \resizebox{.695\textwidth}{.08\textwidth}{ \mytab}}}}
\put(9,79){\textcolor{red}{\textbf{(f)}}}
\end{picture}
    \end{minipage}
    \caption{Q-factor versus launch power for (a) simulation and (b) experiment corresponding to the transmission of a SC-DP 16QAM 34 GBd signal along 17$\times$70km of LEAF. Implementation in the EK-VCK190-G-ED Xilinx FPGA of the (c) biLSTM eq., (d) Deep CNN eq., (e) CDC block. Table (f) shows the latency, hardware usage and throughput for the different designs required to recover 61 symbols simultaneously as well as the estimated amount of FPGAs required to achieve 400G transmission. }
    \label{fig:results}
\end{figure*}
At the transmitter, a DP-16QAM 34~Gbd symbol sequence was mapped out of the data bits generated by a Mersenne Twister algorithm. Then, a digital RRC filter with 0.1 roll-off was applied. The resulting filtered digital samples were resampled and uploaded to a digital-to-analog converter (DAC) operating at 88 GSamples/s. The output of the DAC was amplified by a four-channel electrical amplifier which drove a dual-polarization IQ Mach-Zehnder modulator, modulating the continuous waveform carrier produced by an external
cavity laser at $\lambda = 1.55 \, \mu m$. 
The resulting optical signal was transmitted over 17$\times$70 km spans of LEAF, with the loss in each fiber span being fully compensated at its output using Erbium-doped fiber amplifiers (EDFAs). The EDFAs noise figure was in the 4.5 to 5 dB range. The parameters of the LEAF are: attenuation coefficient $ \alpha = 0.225$ dB/km, chromatic dispersion coefficient $D = 4.2$ ps/(nm$\cdot$km), and effective nonlinear coefficient $\gamma$ = 2 (W$\cdot$ km)$^{-1}$. On the RX side, the optical signal was converted to the electrical domain using an integrated coherent receiver. The resulting signal was sampled at 80~Gsamples/s by a digital sampling oscilloscope and processed using the offline DSP described in~\cite{kuschnerov2010data}. Thereafter, the resulting soft symbols were used as input for the NN equalizers. Finally, the pre-FEC BER was evaluated from the signal at the NN output.
Concerning the simulation, we mimic the experimental transmission setup. The signal propagation along the fiber was simulated by solving the Manakov equations. At the receiver, after full CDC and downsampling to the symbol rate, the received symbols were normalized to the transmitted ones. In addition, Gaussian noise was added to the data signal, representing additional transceiver distortions observed in the experiment. As a result, the Q-factor level of the simulated data matched the experimental one.

Unlike the NN equalizer, which operates with 1 Sa/symbol, the DBP operated with 2.3 Sa/symbol  (and with 1 StpS with the fiber parameters optimized for its best performance). We trained the NN with the data retrieved when setting the highest launch power, namely with $2^{18}$ symbols randomly picked after each epoch from a training dataset of size $2^{20}$, mini-batch size of 2000 and learning rate equal to 0.0005 over 30k epochs\footnote{For applying the trained model to other launch powers, we used transfer learning\cite{freire2021transfer}, thus requiring less than 5 epochs to adjust the NN weights to the other launch powers.}. The test dataset had $2^{18}$ symbols never considered before (not used in the training phase). This dataset was used to report performance in terms of the Q-factor calculated from the BER.

\vspace{-1mm}
\section{Results and Discussion}

Figures~\ref{fig:results} summarize our studies of computational complexity and optical performance. First, the results referring to the simulated data are given in Fig.~\ref{fig:results}~(a). We can observe that the biLSTM equalizer gives approximately the same performance as 1 StpS DBP, improving the optimal power from $-$1 dBm to 2 dBm and the Q-factor by 1.3 dB. The deep CNN performed worse, improving the optimal power from $-$1 dBm to 1dBm and the Q-factor by 0.8~dB. Moving to the experimental data\footnote{The Q-factor produced by the Python model and the FPGA implementation were virtually identical because we did not consider quantization at this moment.}, we observe that biLSTM and deep CNN provided 1.7~dB and 1~dB in Q-factor improvement, respectively, and the optimal power shifts from 0~dBm to 1~dBm in both cases. For the DBP case, the Q-factor improvement in the experiment was 1.5~dB and in the simulation was 1.3~dB. Noticeably, in the experimental study, the NN led to a higher Q-factor improvement, even in the linear regime, when compared with the simulation results\footnote{This highlights the ability of the NN to mitigate the impact of not only the Kerr nonlinearity, but also of the components.}.

Figures~\ref{fig:results} (c), (d), and (e) show the real implementation of the biLSTM, the deep CNN equalizers, and the CDC, respectively, on the state-of-the-art EK-VCK190-G-ED Xilinx FPGA chip\cite{FPGA}. This chip is partitioned into 40 clock regions, with the blue areas in Figs.~\ref{fig:results} (c), (d), and (e) corresponding  to the used chip resources. In Fig.~\ref{fig:results} (f), we summarize the most important information on the implementation of such equalizers in the FPGA. We come to three important conclusions: 1) although biLSTM provided higher Q-factor improvement, its feedback loop connections caused a bottleneck in the design, resulting in higher latency (33 $\mu$s) and lower clock frequency (270 MHz). On the other hand, deep CNN and CDC could be parallelized more efficiently, reducing their latency to 19.9 $\mu$s and 1.1 $\mu$s, respectively. However, because the CDC has one filter whereas the deep CNN has 70 filters, we see this parallelization as being easier in the CDC implementation due to hardware restrictions, resulting in an operating frequency of 524 MHz for the CDC case and 244 MHz for the CNN case. Fig.~\ref{fig:results} (e) clearly shows the CDC parallelization; 2) with respect to FPGA utilization, the biLSTM equalizer was the only one that used Block RAM (BRAM) to save future/past recurrent states, while CNN and CDC did not need such blocks. Concerning the usage of DSP slices, LUT, and FF in each equalizer type, the biLSTM used 64$\%$ DSP slices and 13$\%$ of LUT and FF, the deep CNN used 30$\%$ DSP slices, 13$\%$ of LUT and 21$\%$ of FF, and the CDC used 54$\%$ DSP slices and 1$\%$ of LUT and FF; 3) with regard to throughput, the clock frequency is the maximum that each implementation can handle to comply with a zero-negative slack design. In this sense, the total throughput considering 16QAM modulation format is 66G, 60G, and 127G, respectively, for the biLSTM, the deep CNN, and the CDC block. Finally, considering that the FPGA can safely use 80$\%$ of its resources without having routing/congestion problems and using the maximum utilization reported in Fig.~\ref{fig:results} (f), we observe that 400G transmission can be achieved using an equivalent FPGA that has the same capacity as $\approx$ 5  FPGAs (VCK190) in the case of biLSTM, $\approx$3  FPGAs in the case of deep CNN, and $\approx$ 2  FPGAs in the case of CDC.

\vspace{-1.5mm}
\section{Conclusions}
We demonstrate the FPGA implementation of a recurrent NN equalizer with reduced complexity. We show that the biLSTM equalizer requires only $\approx\!2.5 $ times more FPGA resources than the CDC block. Furthermore, the biLSTM equalizer can outperform DBP with 1StpS, showing its potential to mitigate nonlinear fiber transmission effects. \newline
\footnotesize
\textbf{Acknowledgements:} This work has been supported by the EU H2020  Marie Skodowska-Curie Action project REAL-NET (No. 813144) and EPSRC project TRANSNET.
\normalsize
\linespread{1.0}
\printbibliography

\vspace{-4mm}

\end{document}